\documentclass[a4paper,12pt]{article}
\usepackage[utf8x]{inputenc}

\usepackage{graphicx}
\usepackage{amsmath}
\usepackage{array}

\newcommand{\jweq}[1]{(\ref{#1})}
\newcommand{\emath}{\mathrm{e}}
\newcommand{\D}{\mathrm{d}}

\title{Effect of Partial Absorption on Diffusion with Resetting}
\author{Justin Whitehouse$^{(1)}$, Martin R. Evans$^{(1)}$ and Satya N. Majumdar$^{(2)}$ \\[2ex]
$^{(1)}$ SUPA, School of Physics and Astronomy, University of Edinburgh,\\ Mayfield Road, Edinburgh EH9 3JZ, United Kingdom\\[1ex]
$^{(2)}$ Universit\'{e} Paris-Sud, CNRS, LPTMS, UMR 8626,\\
 Orsay F-01405, France
} 

\date{\today}

\begin{document}

\maketitle

\begin{abstract}
The effect of partial absorption on a diffusive particle which stochastically resets its position with a finite rate $r$ is considered. The particle is absorbed by a target at the origin with absorption `velocity' $a$; as the velocity $a$ approaches $\infty$ the absorption property of the target approaches that of a perfectly-absorbing target. The effect of partial absorption on first-passage time problems is studied, in particular, it is shown that the mean time to absorption (MTA) is increased by an additive term proportional to $1/a$.
The results are extended to multiparticle systems where independent searchers, initially uniformly distributed with a given density, look for a single immobile target. It is found that the average survival probability $P^{av}$ is modified by a multiplicative factor which is a function of $1/a$, whereas the decay rate of the typical survival probability $P^{typ}$ is decreased by an additive term  proportional to $1/a$.

\end{abstract}


\section{Introduction}\label{s:Intro}

Strategies to tackle search problems typically involve a mixture of local steps and long-range moves\cite{Bartumeus2009}.  The local part of the search is often considered as a diffusive process whereas the long-range moves may be drawn from some L\'{e}vy distribution\cite{Oshanin2009} or, as studied recently, may be the stochastic process of `resetting' the search to some preferred position\cite{Evans2011a,Evans2011b,Manrubia1999,Montero2012b}.  From a theoretical perspective the dynamics of the searcher seeking its target may be simply modelled by a diffusive particle or random walker which is absorbed by the target when it arrives at the target's position. Then a measure of the efficiency of the search is the mean time to absorption of the process\cite{Redner2001}.

One simplification inherent in this description is that absorption occurs instantaneously i.e. the searcher instantaneously finds or identifies the target upon contact.  Of course real searches are less reliable and other factors come into play e.g. human or animal error may occur in foraging or searching for a face in a crowd; there
may be stochasticity in the biochemical binding reaction of a protein searching for a target promoter site on a length of DNA.  Moreover, in certain situations the target itself may switch stochastically between states that are visible or invisible to the searcher, for example a fleeing prey hiding from  a predator  or a binding site that switches from being available to unavailable through external factors.

In the reaction kinetics literature a partial reaction is often modelled by a so-called `\emph{radiation boundary condition}' where the probability density flux into the target site is proportional to the probability density at the target site\cite{Redner2001,Ben-Naim1993,Sano1979}. However, here we find it more convenient to instead use a `sink' term in the master equation which represents the loss of probability density from the target site. It has been shown \cite{Wilemski1973,Szabo1984} that the two approaches are equivalent.

Our aim in this paper is to investigate the effects of a stochastic absorption process of the searcher by the target.  We shall consider  the diffusion process with stochastic resetting and introduce a finite rather than infinite absorption rate of the searcher by the target. We will refer to this as partial absorption as opposed to full absorption.  The mathematical problem then becomes, rather than computing the mean first passage time to  a target site, the mean time to absorption by that site.  We investigate the robustness of the results of \cite{Evans2011a,Evans2011b} to partial absorption. We find that generally for high absorption rate the changes are small, so that the results are indeed robust. However, the changes in some properties are more significant than others. For example the mean time to absorption differs from the mean first passage time by an additive constant whereas the survival probability of the target decays with time exponentially with a rate that depends on the absorption rate.

The paper is organised as follows. In section 2 we define the model, discuss boundary conditions relevant to partial absorption and identify important dimensionless quantities in the problem. In section 3 we  compute the survival probability of a single diffusive particle and compute the asymptotics. In section 4 we study the mean time to absorption and how this may be minimised with respect to the resetting rate. In section 5 we consider many independent particles and compute the average and typical survival probabilities for a target at the origin.  We conclude in section 6.

\section{Model}\label{s:Model}

The model considered is that of a diffusive particle which stochastically resets its position and can be absorbed at the origin. To study the probability distribution, $p(x, t|x_0)$, of the particle at time $t$ after having started from initial position $x_0$ we construct the (forward) Master equation
\begin{equation}\label{eq:master}
  \frac{\partial p(x, t|x_0)}{\partial t} = D \frac{ \partial^2 p(x, t|x_0)}{\partial x^2} -rp(x, t|x_0)  +r\delta(x-x_0) -a p(0,t) \delta(x)\;.
\end{equation}
The final term in the Master equation represents the partial absorption process at the origin. The particle is absorbed with rate $a\delta(z)$, where the constant $a$ has dimensions of velocity and we refer to it as the absorption velocity. The first term on the right hand side corresponds to diffusion with diffusion constant $D$,
and the second and third terms correspond to resetting from all points $x$ to the point $x_0$ with rate $r$.

We study the \emph{survival probability} $q(z,t)$ of a diffusive searcher (particle) after time $t$ given that it started at position $z$. 
It is convenient to consider  the backward Master equation for the survival probability  which reads as  follows\cite{Evans2011a}:
\begin{equation}\label{eq:bmaster}
  \frac{\partial q(z,t)}{\partial t} = D \frac{ \partial^2 q(z,t)}{\partial z^2} -rq(z,t) +rq(x_0,t)-aq(0,t)\delta(z)\;.
\end{equation}
The process of resetting to position $x_0$ corresponds to the two terms on the right hand side proportional to the resetting rate $r$, which represent loss from all positions and gain at $x_0$ of probability respectively.
The final term on the right hand side represents 
absorption at the origin.
The system with no absorbing target site (or a target which does not interact with the searcher) is recovered by taking $a = 0$. The total absorption limit can be recovered by taking $a\to\infty$,  as  we now discuss.

\subsection{Boundary Conditions}\label{s:BC}

The most commonly studied boundary condition in the diffusive  process literature is the \emph{totally absorbing boundary condition} wherein, once the particle meets the boundary, it is certain that the particle is absorbed  and  leaves the system instantaneously. The totally absorbing boundary condition is therefore defined, in terms of the probability density $p(x,t)$ of the particle at the boundary $x_B$, as $p(x_B,t) = 0$.

To describe a system which includes a `reflecting wall', it is natural to use what is commonly referred to as the `\emph{reflecting boundary condition'}. Here, because all particles which try to reach the boundary location are `reflected' back away from the boundary, the net flux into the boundary site is zero. This means that this boundary condition can be straightforwardly expressed as 
\begin{equation}\label{eq:ref_bc_defn}
  \frac{\partial p(x,t)}{\partial x}\bigg|_{x=x_B} = 0\;.
\end{equation}


The so-called radiation boundary condition \cite{Sano1979,Redner2001} is  given by
\begin{equation}\label{eq:rad_bc_e.g.}
  \frac{\partial p(x,t)}{\partial x}\bigg|_{x=x_B} = \frac{a}{D}p(x_B, t)\;.
\end{equation}
This can be interpreted as a boundary which displays the characteristics of both the totally absorbing and reflecting cases i.e. there is partial absorption.

In principle the diffusion equation can be solved with the radiation boundary condition by using the appropriate Green function, but it may be that this Green function is difficult to find. It then may be more convenient to use what is known as a `\emph{sink term}' in the Master equation instead. In general terms, a sink term represents the loss of probability density from a region of space as a consequence of absorption, attenuation, or some other similarly natured process.

To see the equivalence  explicitly \cite{Wilemski1973}
we consider an absorbing region of width $2x_B$ centered at the origin
of the real  line, at the {\em boundaries} of which an incident diffusive particle is  absorbed with rate
$a$.
The Master equation for the probability density of the particle in this system reads
\begin{equation}\label{eq:bc_eg_Master}
 \frac{\partial p(x,t)}{\partial t} = \frac{\partial^2 p(x,t)}{\partial x^2} - a\delta(x_B - |x|)p(x,t)\;.
\end{equation}
 We define the survival probability of the particle at time $t$ as 
the integral over all space
\begin{equation}\label{eq:bc_eg_n}
  n(t) = \int_{-\infty}^{+\infty}p(x,t) \D x 
\end{equation}
and  probability that the particle is at a position $|x|>x_B +\epsilon$ as
\begin{equation}\label{eq:bc_eg_m}
  m(t) = \int_{-\infty}^{-(x_B+\epsilon)} p(x,t) \D x + \int^{\infty}_{x_B+\epsilon} p(x,t) \D x \;.
\end{equation}
We assume that far away from the absorbing region that 
\begin{equation}\label{qe:bc_eg_grad_p_infty}
  \frac{\partial p(x, t)}{\partial x}\bigg|_{x\to\pm\infty} = 0
\end{equation}
and perform the necessary integrals on (\ref{eq:bc_eg_Master}) to obtain
\begin{equation}
  \frac{\partial n(t)}{\partial t} = -a [ p(-x_B, t) + p(x_B, t) ] 
\end{equation}
from (\ref{eq:bc_eg_n}), and
\begin{equation}
  \frac{\partial m(t)}{\partial t} = D \left[ \frac{\partial p}{\partial x}\bigg|_{x=-x_B} - \frac{\partial p}{\partial x}\bigg|_{x=x_B}  \right]
\end{equation}
from (\ref{eq:bc_eg_m}). Now we demand that
\begin{equation}\label{eq:bc_eg_demand_mn}
  \lim_{\epsilon\to0}\frac{\partial m}{\partial t} = \frac{\partial n}{\partial t}
\end{equation}
to find
\begin{equation}
  \frac{\partial p}{\partial x}\bigg|_{x=\pm x_B} = \pm \frac{a}{D} p(\pm x_B,t) \;,
\end{equation}
which is  as the radiation boundary condition quoted above.
Thus we have  shown that a sink term of the form
\begin{equation}\label{eq:sink_e.g.}
  - a \delta(x-x_B) p(x,t)
\end{equation}
in a Master equation of the form of (\ref{eq:master}) is equivalent to the boundary condition (\ref{eq:rad_bc_e.g.}).

As is comprehensively discussed in \cite{Redner2001}, the length $D/a$ is equivalent to the attenuation length in a composite medium. This is a medium in which there is a boundary at $x=0$, and in the region $x>0$ particles can undergo free diffusion, and for $x<0$ the density of particles is attenuated\cite{Ben-Naim1993}. 
It can be shown for this system that the probability density within the attenuating region decreases linearly to 0 at the depth $x = -D/a = -l$, hence we refer to $l$ as the attenuation depth.
This furnishes a physical interpretation of the  characteristic length scale
$D/a$.

\subsection{Dimensionless variables}\label{s:DimVars}
We first define the inverse length scale $\alpha_0$
\begin{equation}
  \alpha_0 = \left( \frac{r}{D}\right)^{1/2} \;,
\label{alf0def}
\end{equation}
where $\alpha_0^{-1}$ represents the characteristic diffusion length between resetting events.

For convenience we summarise here some dimensionless combinations of parameters which will appear later:
\begin{eqnarray}
  \theta &=& \sqrt{\frac{r}{D}}|x_0| = \alpha_0 |x_0| \label{eq:theta_defn} \\
  \phi_0 &=& \frac{\sqrt{rD}}{a} = \frac{\alpha_0 D}{a}\;.\label{eq:phi_0_defn}
\end{eqnarray}

The dimensionless quantity $\theta$ measures the ratio of the distance of the 
resetting position $x_0$ from the target (in this case at the origin) to the inverse length $\alpha_0$. The dimensionless quantity $\phi_0$ measures the ratio of the attenuation length $D/a$ to the characteristic diffusion length $1/\alpha_0$.

\section{Survival Probability Calculation}\label{s:SurvivalProb}

To solve the Master  equation (\ref{eq:bmaster}) for the survival probability
with partial absorption we perform the Laplace transform on the variable $t$. We define the Laplace Transform of the survival probability as
\begin{equation}\label{eq:LT_defn}
  \tilde{q}(z,s) = \int_0^\infty q(z,t)e^{-st} \D t \; .
\end{equation}
We use this definition and the initial condition $q(z,0) = 1$ to take  the Laplace Transform of  equation (\ref{eq:bmaster}):
\begin{equation}\label{eq:laplace_master}
D\frac{\partial^2 \tilde{q}(z,s)}{\partial z^2} - (r+s)\tilde{q}(z,s) = -1 - r\tilde{q}(x_0,s) + a\tilde{q}(0,s)\delta(z) \;.
\end{equation}

We first find the solution to the homogeneous equation
\begin{equation}
  D\frac{\partial^2 \tilde{q}(z,s)}{\partial z^2} - (r+s)\tilde{q}(z,s) = 0 \;,
\end{equation}
which yields
\begin{equation}
  \tilde{q}(z,s) = Ae^{\alpha z} + Be^{-\alpha z}
\end{equation}
where
\begin{equation}
  \alpha(s) = \left( \frac{r+s}{D}\right)^{1/2}\;.
\end{equation}
(In particular from (\ref{alf0def})  we have   $\alpha_0 = \alpha(0)$.)
From the full equation (\ref{eq:laplace_master}) (for $z\neq 0$)
 we find the particular solution
\begin{equation}\label{eq:lap_part_sol}
  \tilde{q}(z,s) = Ae^{\alpha z} + Be^{-\alpha z} + \frac{1 + r\tilde{q}(x_0,s)}{r+s}\;.
\end{equation}
We require that our solution be finite in the limits $z\to\pm\infty$, which 
implies separate solutions $\tilde{q}(z,s)_+$ and $\tilde{q}(z,s)_-$ for the regions where $z$ is greater than and less than zero respectively. Using the required continuity at $z=0$ to find that $A=B$, we obtain the result
\begin{equation}\label{eq:q_tilde_pm}
  \tilde{q}_\pm(z,s) = Ae^{\mp\alpha z} + \frac{1 + r\tilde{q}(x_0,s)}{r+s} \;.
\end{equation}



\begin{figure}[h]
 \centering
 \includegraphics[width=0.5\textwidth]{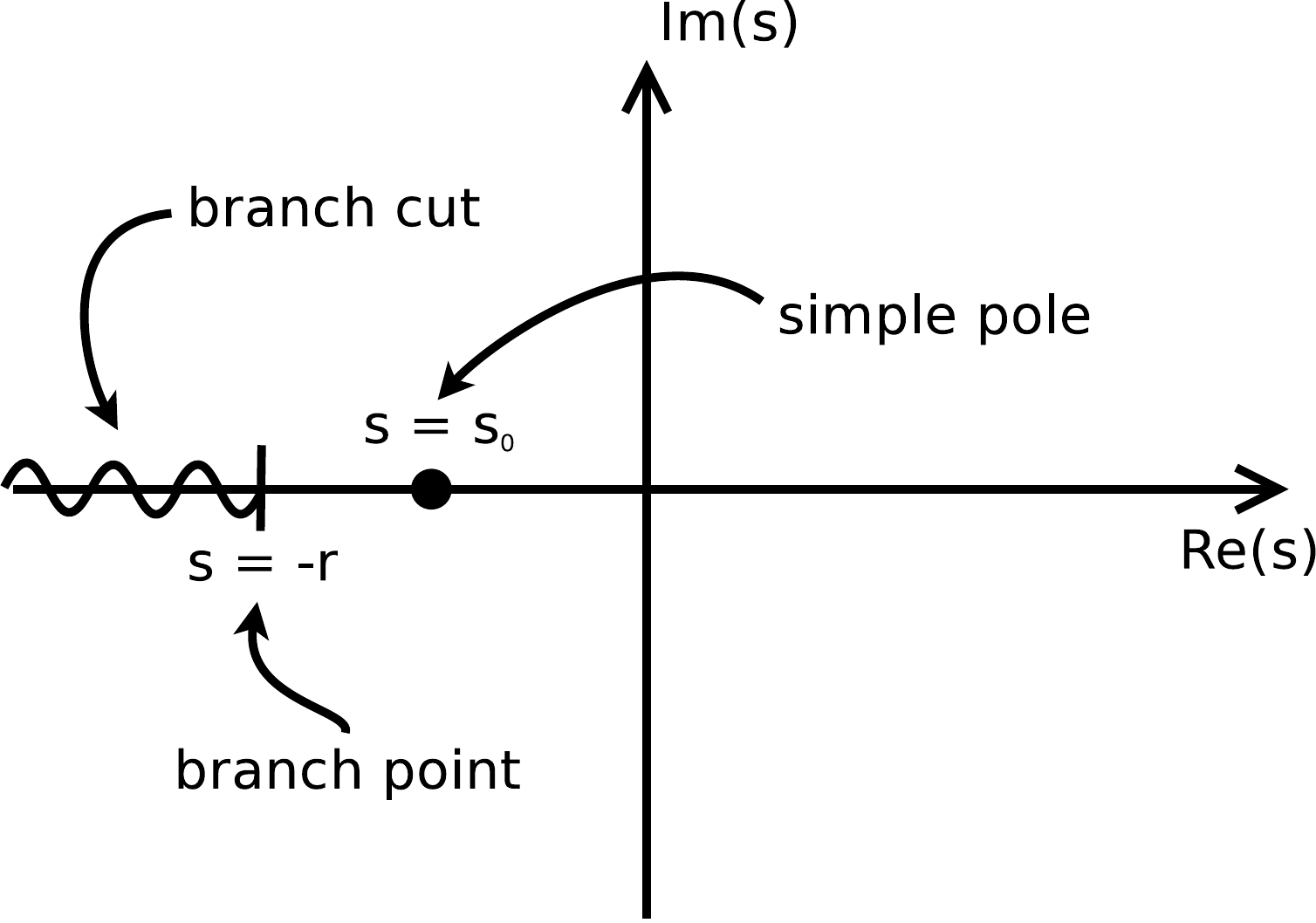}
 \caption{Sketch of the singularities of the function (\ref{eq:lapq_finalfinal}) in the complex $s$ plane.}
  \label{fig:complex_plane}
\end{figure}

Now we can use the discontinuity in the derivative which comes from the delta function in equation (\ref{eq:bmaster}) to find $A$: 
\begin{equation}\label{eq:int_lap_mstr}
  D\left[\frac{\partial \tilde{q}_+}{\partial z}- \frac{\partial \tilde{q}_-}{\partial z}\right]_{z=0}   =   a\tilde{q}(0,s) \;.
\end{equation}
From (\ref{eq:q_tilde_pm}) and (\ref{eq:int_lap_mstr}) we derive the solution

\begin{equation}\label{eq:lapq_finalfinal}
 \tilde{q}(z,s) = \frac{1 - \tilde{Q}(z,s)}{s + r\tilde{Q}(x_0,s)}
\end{equation}
where
\begin{equation}
  \tilde{Q}(z,s) = \frac{e^{-\alpha |z|}}{1 + 2\phi(s)}\;,
\end{equation}
$\phi(s)$ is given by
\begin{equation}\label{eq:phi_s_defn}
  \phi(s) = \frac{\sqrt{(r+s)D}}{a}\;,
\end{equation}
and $\phi(0) = \phi_0$ as given in \jweq{eq:phi_0_defn}.
The long time behaviour of the survival probability can be determined from equation (\ref{eq:lapq_finalfinal}). As illustrated in Fig. \ref{fig:complex_plane}, the analytic structure of this function in the complex $s$ plane is a branch point $s = -r$, and a simple pole at $s=s_0$ which satisfies:
\begin{equation}\label{eq:poles_satisfy}
  s_0(2\phi(s_0)+1) + re^{-\sqrt{\frac{(r+s_0)}{D}}|z|} = 0
\end{equation}
where
\begin{equation}
 -r < s_0 \le 0\;.
\end{equation}

In the large time limit, the  residue from $s_0$ dominates the solution to the Bromwich inversion formula for this function.
For convenience, we make a change of variable
\begin{equation}
  s_0 = r(u_0 - 1) \;.
\end{equation}
Under this substitution equation (\ref{eq:poles_satisfy}) becomes
\begin{equation}\label{eq:poles_u_0}
  (u_0-1)(2\phi_0\sqrt{u_0} + 1) + \exp\left(-\sqrt{\frac{ru_0}{D}}|z|\right) = 0 \;, 
\end{equation}
where $\phi_0$ is given in \jweq{eq:phi_0_defn}. For further convenience we consider the case where $z = x_0$  (i.e. the initial position coincides with the resetting position) to find
\begin{equation}\label{eq:q_inv_full}
  q(x_0,t) \simeq  \frac{2u_0^{3/2}[1 + \phi_0u_0^{1/2}] e^{r(u_0-1)t} }{(3u_0 - 1)\phi_0 + \theta\phi_0u_0^{1/2}(u_0 - 1) + 2u_0^{1/2} + \theta(u_0 - 1)} \;.
\end{equation}
The important point to note is that the survival probability decays exponentially and
the rate of decay increases with   absorption velocity $a$.
If one expands to leading order in $1/a$ one finds
\begin{equation}\label{eq:q_decay_const}
u_0 \simeq  u^* + \frac{4u^*(1-u^*)}{2(u^*)^{1/2} - \theta (1-u^*)} \, \frac{\alpha_0D}{a} \;,
\end{equation}
where  $u^*$ is the solution of \jweq{eq:poles_u_0} as $a \to \infty$.

\section{Mean Time to Absorption}\label{s:MFPT}

The Mean Time to Absorption (MTA) $T(z)$ for a particle which originated at $z$ is \cite{Redner2001}
\begin{equation}\label{eq:MFPT_defn}
  T(z) =  -\int^{\infty}_{0} t \frac{\partial q(z,t)}{\partial t}\D t
=
\int^{\infty}_{0}  q(z,t)\D t = \tilde{q}(z,s=0).
\end{equation}

In our analysis we are considering mean first time to \emph{absorption} of the searcher by the target, not just the mean first time to coincidence as in the case where the absorption is perfect. This means that it is possible for the searcher to pass \emph{through} the target site without interacting with it. The  MTA as defined above in \jweq{eq:MFPT_defn} is still the appropriate measure of this process. Using equation \jweq{eq:lapq_finalfinal} and the definition in equation \jweq{eq:MFPT_defn} we can write down an expression for the MTA:
\begin{equation}\label{eq:MFPT_2}
  T(z) = \frac{e^{\alpha_0|x_0|}}{r} \left[ 2\phi_0 + 1 - e^{-\alpha_0|z|} \right] \;,
\end{equation}
and in particular
\begin{equation} \label{eq:MFPT_3}
 T(x_0) = \frac{1}{r}\left[ e^{\alpha_0|x_0|} -1\right] + \frac{2e^{\alpha_0|x_0|}}{r}\phi_0  \;.
\end{equation}
Thus the effect of partial absorption is to increase the mean time to absorption through the second term in \jweq{eq:MFPT_3}. To understand better this term it is instructive to consider a more general problem which involves a  resetting \emph{distribution}\cite{Evans2011a} ${\cal P}(x)$, i.e. the particle resets with rate $r$ to a random  position drawn from the distribution ${\cal P}(x)$. The resetting term $rq(x_0,t)$ in the backward Master equation \jweq{eq:bmaster} becomes $r\int dx {\cal P}(x) q(x,t)$ so, for the survival probability $q(z,t)$, the Master equation itself reads
\begin{equation}\label{eq:master_rd}
  \frac{\partial q(z,t)}{\partial t} = D\frac{\partial^2 q(z,t)}{\partial z^2} - rq(z,t) + r\int {\cal P}(x)q(x,t)\D x - aq(0,t)\delta(z) \;.
\end{equation}
The calculation of the mean time to absorption is a straightforward generalisation of that presented above and the result is
\begin{equation}\label{MFPT_rd_0_3}
 T(z) = \frac{1}{2\sqrt{rD}p^*(0)} \left(1 - e^{-\alpha_0|z|} + 2\phi_0 \right) \;,
\end{equation}
where
\begin{equation}
 p^*(x) =  \frac{\alpha_0}{2} \int \D x' {\cal P}(x') e^{-\alpha_0|x-x'|} \;.
\end{equation}
$p^*(x)$ is the stationary distribution of the diffusive process with resetting but without any absorption\cite{Evans2011a}. Thus \jweq{MFPT_rd_0_3} tells us that the effect of partial absorption is to increase  the mean time to absorption by an additional term proportional to $1/a$. In fact the precise form of the additional term is simply $1/a p^*(0)$, where $p^*(0)$ is the stationary probability of the searcher  being at the origin (target site) in the presence of resetting but absence of absorption.

\subsection{Minimisation of the MTA with respect to resetting rate $r$}\label{s:MinMFPT}
\begin{figure}[h!]
  \centering
  \includegraphics[width=\textwidth]{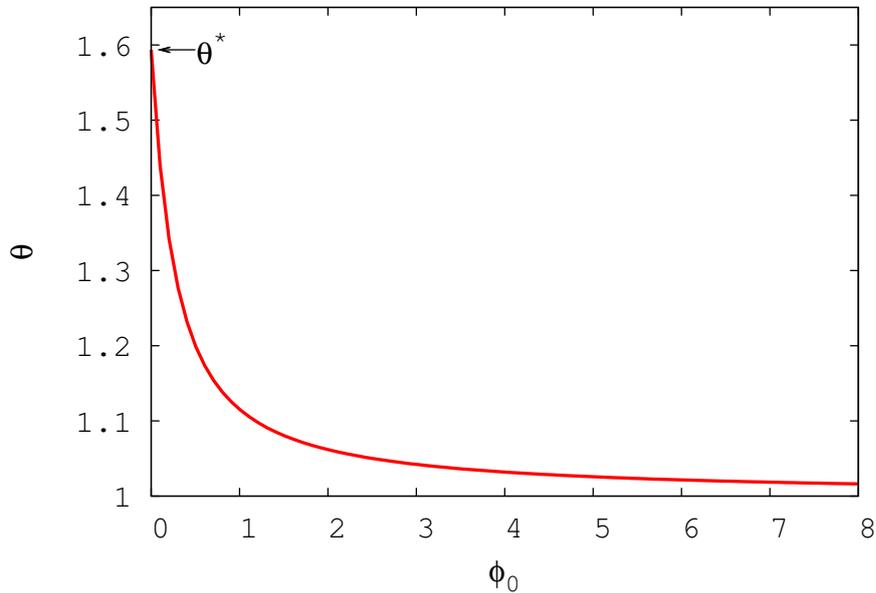}
  \caption{Plot of the value of $\theta$
that  minimises the MTA for a given $\phi_0$. The value of $\theta$ at $\phi_0 = 0$, found numerically to be $\theta^* = 1.5936$, is indicated.}
  \label{fig:minimise_MFPT}
\end{figure}

Returning to \jweq{eq:MFPT_3} we are interested in finding the choice of $r$ which minimises $T(z)$. We start by considering the minimum of $T(x_0)$ with respect to $r$.
Setting $\displaystyle   \frac{\partial T(x_0)}{\partial r} =  0$
yields
\begin{equation}\label{eq:roots_of_1}
 \emath^{-\theta} = 1 + \phi_0-\frac{\theta}{2}- \phi_0\theta \;,
\end{equation}
where $\theta$ and $\phi_0$ are defined in \jweq{eq:theta_defn} and \jweq{eq:phi_0_defn}. The solutions of \jweq{eq:roots_of_1} are shown in Fig. \ref{fig:minimise_MFPT}.

In the following we analyse the solution of \jweq{eq:roots_of_1} in the regimes $a \gg \alpha_0D$ and $a \ll \alpha_0D$ corresponding to strong and weak
absorption.

\begin{figure}[h!]
  \centering
  \includegraphics[width=\textwidth]{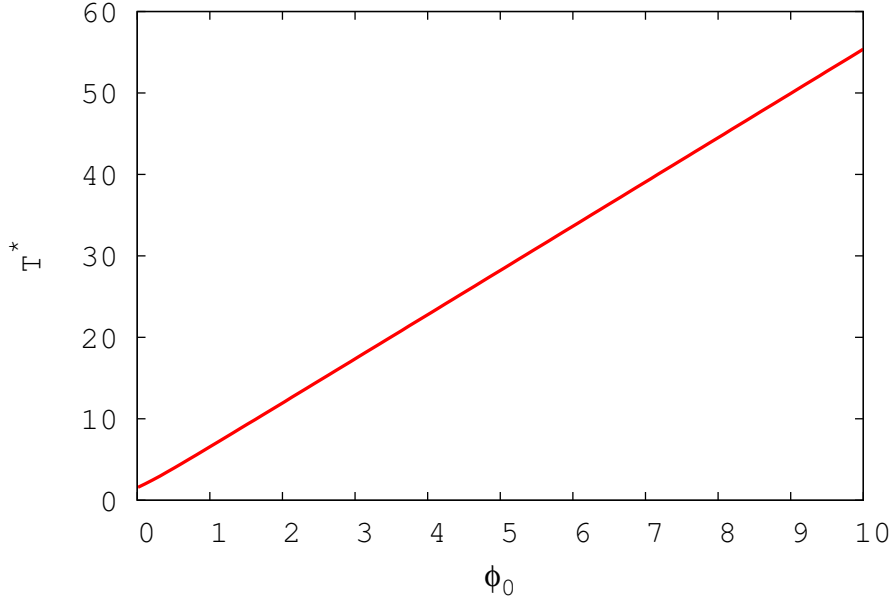}
  \caption{The minimised MTA, $T^*$, as a function of $\phi_0$, in units of $x_0^2/D$. $T^*$ is approximately linear for both small and large $\phi_0$. (See \jweq{eq:small_phi0_T*} and \jweq{eq:large_phi0_T*}.)}
  \label{fig:minimise_MFPT2}
\end{figure}

First we note that for $\phi_0 \ll 1$ (i.e. strong absorption, $a \gg \alpha_0D$) Eq. \jweq{eq:roots_of_1} reduces to the transcendental equation found in the study of total absorption\cite{Evans2011b}:
\begin{equation}
   \frac{\theta^*}{2} = 1 - \emath^{-\theta^*} \;.
\end{equation}
We then find that, to second order in $\phi_0$,
\begin{equation}\label{eq:theta_phi_correction}
  \theta = \theta^* - 2\phi_0 + \frac{2\theta^*}{\theta^*-1}\phi_0^2
\end{equation}
satisfies \jweq{eq:roots_of_1}. It is helpful to rewrite  Eq. \jweq{eq:MFPT_3} as
\begin{equation}
  T(x_0) = \frac{x_0^2}{D\theta^2} \left[ (2\phi_0 + 1)\emath^{\theta} - 1 \right] \;,
\end{equation}
and then, using Eq. \jweq{eq:theta_phi_correction}, it can be shown that for $\phi_0\ll1$ the MTA minimised with respect to $r$ is 
\begin{equation}\label{eq:small_phi0_T*}
  T^*(x_0) \simeq \frac{x_0^2}{D}\,\frac{1}{\theta^*(2-\theta^*)} 
\left[1 + \frac{4}{\theta^*}\phi_0 + \frac{4(3-\theta^*)}{(\theta^*)^2} \phi_0^2 \ldots \right]
\;.
\end{equation}
To study the regime $\phi_0 \gg 1$ (i.e. weak absorption, $a \ll \alpha_0D$), we first rewrite Eq. \jweq{eq:roots_of_1} as
\begin{equation}\label{eq:roots_of_2}
  1 - \theta = \frac{1}{\phi_0} \left( \emath^{-\theta} - 1 + \frac{\theta}{2} \right) \;.
\end{equation}
Then we can see that for $\phi_0 \gg 1$, $\theta\to1^+$, and explicitly 
to first order in $1/\phi_0$
\begin{equation}\label{eq:theta_correction}
  \theta \simeq   1 + \frac{1 - 2\emath^{-1}}{2\phi_0} \;.
\end{equation}
Using this result we find that the minimised MTA is, to leading order in $1/\phi_0$, 
\begin{equation}\label{eq:large_phi0_T*}
  T^*(x_0) \simeq  2 \emath\, \frac{x_0^2}{D}\, \phi_0\;.
\end{equation}
We plot $T^*$ as a function of $\phi_0$ in Fig. \ref{fig:minimise_MFPT2}.

\section{Many Independent Searchers}\label{s:ManySearchers}
We now consider  the multiparticle version of the search process
comprising  a single immobile target
at the origin and many  searchers which are initially uniformly distributed
on the line with uniform density $\rho$.
The searchers are independent of 
each other and the position of each searcher evolves stochastically 
starting at its own initial position to which it resets.
 
The survival probability of the target, $P_s(t)$, is given\cite{Evans2011b} by
\begin{equation}
 P_s(t) = \prod_{i=1}^Nq(x_i,t) \;,
\end{equation}
where $q(x_i, t)$ is the survival probability in the single searcher problem. The positions $x_i$ are independent and distributed uniformly within the box [-L/2, L/2]. The average probability $P_s^{av}(t) = \langle P_s(t)\rangle_x$ and the typical probability $P_s^{typ}(t) = \exp[ \langle \ln P_s(t) \rangle_x ]$, where $\langle\cdot \rangle_x$ denotes averages over $x_i$'s. 

In random additive processes the average of the random variable and its typical or most probable value exhibit the same behaviour. For multiplicative processes (where a product of random variables is considered such as is described here), there exist extreme events which, although exponentially rare, are exponentially different from the typical value of the product\cite{Redner1990}. Thus, for multiplicative processes, it is important to consider both average and typical values.

\subsection{Average Survival Probability of the Target}\label{s:ManyAvgeSurvival}

The average survival probability of the target, $P^{av}_s(t)$, is given by
\begin{equation}
  P^{av}_s(t) = \langle q(x,t) \rangle^N_x = \exp(N \ln[1 - \langle 1 - q \rangle_x ] ) \;,
\end{equation}
where
\begin{equation}
  \langle 1 - q \rangle_x = \frac{1}{L}\int^{L/2}_{-L/2} \D x [1-q(x,t)] \;.
\end{equation}
Taking $N, L \to\infty$ while keeping the density of searchers $\rho = N/L$ fixed (using $q(x,t) = q(-x,t)$) we find:
\begin{equation}
  P^{av}_s(t) \to \exp \left(-2 \rho \int_0^{\infty} \D x[1-q(x,t)] \right) 
\equiv \exp(-2\rho M(t)) \;.
\end{equation}
Defining

\begin{equation}
  \tilde{M}(s) = \int_0^\infty M(t) e^{-st} \D t \;,
\end{equation}
we can use equation \jweq{eq:lapq_finalfinal} to find
\begin{equation}\label{eq:M_s}
  \tilde{M}(s) = \frac{r+s}{sr\alpha}\ln \left( 1 + \frac{r}{s(2\phi(s) + 1)} \right) \;.
\end{equation}
In principle this may be inverted to find $P^{av}_s(t)$. Rather than present the formula which takes the form of a double convolution it is more instructive to examine the effect of partial absorption on asymptotic behaviour directly from \jweq{eq:M_s}.

For convenience we first write
\begin{equation}
  \phi(s) = k(r+s)^{\frac{1}{2}}\mbox{,\quad where }k = \frac{\sqrt{D}}{a} \;,
\end{equation}
so that we can then expand \jweq{eq:M_s} to leading order to find
\begin{equation}\label{eq:small_s_M_leading}
  \tilde{M}(s) = \sqrt{\frac{D}{r}} \frac{1}{s} \left( -\ln s + \ln r -\ln [2k\sqrt{r} + 1] + O(s) \right) \;,
\end{equation}
and using the identity\cite{Roberts1966}
\begin{equation}\label{eq:invLap_Euler}
  \mathcal{L}^{-1} \left[ -\frac{\ln s}{s} \right] = \ln t + \gamma \;,
\end{equation}
where $\gamma$ is Euler's constant, we find then that for long times
\begin{equation}\label{eq:long_t_P_av_phi}
 P^{av}(t) \sim \exp \left[ -2\rho \left( \ln rt + \gamma -  \ln \left( 2\phi_0 + 1 \right) \right) \right] \;.
\end{equation}
Thus the effect of partial absorption is to change the asymptotic decay of the average probability by a multiplicative factor
$(1+2\phi_0)^{2\rho}$.

\subsection{Typical Survival Probability}\label{s:ManyTypSurvival}

The typical survival probability of the target, $P^{typ}_s(t)$, can be expressed as
\begin{equation}\label{eq:Ptyp}
  P^{typ}_s(t) = \exp \sum^N_{i=1}\langle \ln q(x_i, t) \rangle_x = \exp \left[2\rho\int_0^{L/2} \D x\ln q(x,t) \right] \;.
\end{equation}
In the long-time limit we can use Eq. \jweq{eq:q_inv_full} to show that
\begin{equation}\label{eq:Ptyp_exp}
  \int_0^\infty \D x_0 \ln q(x_0, t) \simeq const - t\int_0^\infty \D x_0|s_0(x_0)| 
\end{equation}
and $P^{typ}(t)\sim \exp(-2I\rho t)$. 

We can calculate the integral $I = \int_0^\infty dx_0s_0(x_0)$ using the following procedure.
First, using equation \jweq{eq:poles_satisfy} we make the substitution $z = -s_0/r$ to find that $z$ satisfies
\begin{equation}\label{eq:z_solves}
  z\left(\phi_0(1-z)^{1/2}+1\right) = \exp(-\theta(1-z)^{1/2}) \;,
\end{equation}
which we use to rewrite $I$, by eliminating $\theta$, as an integral over $dz$:
\begin{equation}
  I = \sqrt{rD}\int_{z_0}^0 \D z \left( \frac{\ln z }{(1-z)^{1/2}} + \frac{\ln[(\phi_0(1-z)^{1/2} + 1)]}{(1-z)^{1/2}}\right) \;,
\end{equation}
where
\begin{equation}
  z_0 = -\frac{1}{2\phi_0^2} \left( 1 - \sqrt{1 + 4\phi_0^{2}} \right) \;.
\end{equation}
We make the substitution 
\begin{equation}\label{eq:subs_y}
y^2 = 1-z
\end{equation}
to obtain
\begin{equation}\label{eq:I}
  I = 2\sqrt{rD}\int_1^{y_0} \D y \left( \ln(1-y^2) + \ln[(\phi_0y + 1)]\right) \;,
\quad\mbox{where }y_0^2 = 1 - z_0 \;. 
\end{equation}
The first term in this integral, which will henceforth be labelled $I_0$,
\begin{equation}
  I_0 = 2\sqrt{rD}\int^{y_0}_1 \D y \left[ \ln(1-y) + \ln(1+y) \right] 
\end{equation}
which can be evaluated to obtain
\begin{equation}\label{eq:I0_final}
  I_0 = 4\sqrt{rD} \left[ 1 - \ln2 - y_0 + \frac{(1+y_0)}{2}\ln(1+y_0) - \frac{(1-y_0)}{2}\ln(1-y_0) \right] \;.
\end{equation}
For $\phi_0 \ll 1$ we find
\begin{equation}\label{eq:lim_y0^2}
  y_0^2 = \left[ 1 + \frac{1}{2\phi_0^2} \left( 1 - \sqrt{1 + 4\phi_0^2} \right) \right] \simeq \phi_0^2
\end{equation}
and consequently
\begin{equation}\label{eq:lim_I0}
  I_0 \simeq 4\sqrt{rD}\left( 1 - \ln2 - \frac{\phi_0^3}{6} + O(\phi_0^4) \right) \;.
\end{equation}

To calculate the second term, $I_1$, in \jweq{eq:I} we again use the substitution \jweq{eq:subs_y} to find
\begin{equation}\label{eq:I_1_soln}
  I_1 = -\frac{2\sqrt{rD}}{\phi_0} \left[ \left( \phi_0 + 1 \right) \ln \left( \phi_0 + 1 \right) - \left( y_0\phi_0 + 1 \right) \ln \left( \phi_0 y_0 + 1 \right) - \phi_0 \left( 1-y_0 \right) \right] \;.
\end{equation}
From \jweq{eq:lim_y0^2} we see that, for $\phi_0 \ll 1$, 
\begin{equation}\label{eq:lim_Ip}
  I_1  \simeq  -4\sqrt{rD} \left[ \frac{\phi_0}{4} + O(\phi_0^2) \right] \;.
\end{equation}

Combining the results \jweq{eq:lim_I0} and \jweq{eq:lim_Ip} we find that for $\phi_0 \ll 1$
\begin{equation}\label{eq:lim_I}
  I \simeq 4\sqrt{rD} \left( 1 - \ln2 - \frac{\phi_0}{4} \right) \;.
\end{equation}
In this limit, from \jweq{eq:lim_I0} and \jweq{eq:lim_Ip}, the correction to the total absorption decay rate is linear in $\phi_0$: 
\begin{equation}
  P^{typ}(t) \sim \exp \left[-8\sqrt{rD}\left(1 - \ln2 - \frac{\phi_0}{4} \right) \rho t \right] \;.
\end{equation}

The full result for $P^{typ}(t)$ is
\begin{eqnarray}\label{eq:Ptyp_sol_long_array}
  P^{typ}(t)	& \sim 	& \exp \left\{ -8\rho t\sqrt{rD} \left[  1 - \ln2 - y_0 + \frac{(1+y_0)}{2}\ln(1+y_0) - \frac{(1-y_0)}{2}\ln(1-y_0) \right. \right. \nonumber\\
		& 	& - \left. \left.
 \left( \frac{\phi_0 + 1}{2\phi_0} \right) \ln \left( \phi_0 + 1 \right) - \left( \frac{y_0\phi_0 + 1}{2\phi_0} \right) \ln \left( \phi_0 y_0 + 1 \right) - \left( \frac{1-y_0}{2} \right) \right] \right\}
\end{eqnarray}

\section{Conclusion}

In this work we have considered the dynamics of a diffusive searcher in a system with a partially absorbing target, as defined in \jweq{eq:master}, which gives a more realistic description of many, varied search processes. Our study has revealed some straightforward, significant and intuitive consequences for the dynamics as a direct result of imperfection in the absorption at the target.

We see from Section \ref{s:SurvivalProb}, and equations \jweq{eq:q_inv_full} and \jweq{eq:q_decay_const} in particular, that the survival probability of the searcher (or target) decreases exponentially with time, with a decay constant which increases as the absorption constant $a$ increases. As the target comes closer to being perfectly absorbing, the survival probability decays away much sooner.

Our study of the MTA in Section \ref{s:MFPT} has revealed that the mean time to absorption is increased by an additive term 
 $1/a p^*(0)$ (see (\ref{MFPT_rd_0_3})). As $a$ decreases and the target is made more `imperfect' the mean time to absorption increases, as expected.

For multiplicative processes, such as the many searcher problem analysed in Section \ref{s:ManySearchers}, the distinction between \emph{typical} and \emph{average} probabilities becomes significant. This is emphasised by the difference in the form of effects of imperfect absorption on the typical and average survival probabilities of the target in the many searcher system. We see from Section \ref{s:ManyAvgeSurvival} that $P^{av}(t)$ is modified by multiplicative factor $(1+2\phi_0)^{2\rho}$, whereas in Section \ref{s:ManyTypSurvival} it is the decay rate of $P^{typ}(t)$ which is decreased by factor proportional to $1/a$. 

An important quantity that emerges from this this work is the dimensionless ratio $\phi_0 = \alpha_0D/a$. This quantity is a ratio of length scales characteristic to the system: the length scale $1/\alpha_0 = \sqrt{D/r}$ has already been established as the characteristic displacement of the searcher due to diffusion between reset events\cite{Evans2011a}; the length $D/a$ is the attenuation depth in the composite medium discussed in \cite{Redner2001}. Thus the dimensionless variable $\phi_0$ represents the ratio of these two length scales, and the dynamics are only modified in terms of this ratio and not the absolute strength of the imperfection which has been introduced to the target. It is the competition between resetting and the absorption that 
controls the dynamics of the system.

In future work it would be of interest to extend the present calculations to higher dimensions.
Also a related problem of interest \cite{Oshanin2009}  would be to consider discrete time jump processes where the searching particle 
may jump over the target  thus reducing the effective absorption rate.\\[2ex]

\noindent{\bf Acknowledgements:}
JW  acknowledges financial support from EPSRC (UK) via the Scottish Doctoral Training Centre in Condensed Matter Physics.

\appendix


\bibliography{EPADR_paper}{}
\bibliographystyle{revtex}


\end{document}